\documentclass[aps,prb,reprint,twocolumn,superscriptaddress,amsmath,amssymb]{revtex4}
\usepackage{graphicx}%
\usepackage{float}
\usepackage[T1]{fontenc}
\usepackage{xcolor}
\usepackage{ulem}
\usepackage[colorlinks=true, urlcolor=blue, linkcolor=blue, citecolor=blue, bookmarks, pdftitle={article}]{hyperref}
\usepackage[]{units}
\usepackage{times}
\usepackage{bm}

\newcommand{\comment}[1]{}

\renewcommand{\emph}{\textit}

\begin{document}
\title{Interlayer exciton valley polarization dynamics in large magnetic fields}

\author{Johannes Holler}
\affiliation{Institut f\"ur Experimentelle und Angewandte Physik,
	Universit\"at Regensburg, 93040 Regensburg, Germany}

\author{Malte Selig}
\affiliation{Institut f\"ur Theoretische Physik, Technische Universit\"at Berlin, 10587 Berlin, Germany}

\author{Michael Kempf}
\affiliation{Institut f\"ur  Physik, Universit\"at Rostock, 18059 Rostock, Germany}	

\author{Jonas Zipfel}
\affiliation{Institut f\"ur Experimentelle und Angewandte Physik, Universit\"at Regensburg, 93040 Regensburg, Germany}
\affiliation{Molecular Foundry, Lawrence Berkeley National Laboratory, Berkeley, California 94720, USA}

\author{Philipp Nagler}
\affiliation{Institut f\"ur Experimentelle und Angewandte Physik, Universit\"at Regensburg, 93040 Regensburg, Germany}

\author{Manuel Katzer}
\affiliation{Institut f\"ur Theoretische Physik, Technische Universit\"at Berlin, 10587 Berlin, Germany}

\author{Florian Katsch}
\affiliation{Institut f\"ur Theoretische Physik, Technische Universit\"at Berlin, 10587 Berlin, Germany}

\author{Mariana V. Ballottin}
\affiliation{High Field Magnet Laboratory (HFML - EMFL), Radboud University, 6525 ED Nijmegen, The Netherlands}

\author{Anatolie A. Mitioglu}
\affiliation{High Field Magnet Laboratory (HFML - EMFL), Radboud University, 6525 ED Nijmegen, The Netherlands}

\author{Alexey Chernikov}
\affiliation{Institut f\"ur Experimentelle und Angewandte Physik,
	Universit\"at Regensburg, 93040 Regensburg, Germany}
\affiliation{Dresden Integrated Center for Applied Physics and Photonic Materials (IAPP) and W\"urzburg-Dresden Cluster of Excellence ct.qmat, Technische Universit\"at Dresden, 01062 Dresden, Germany}

\author{Peter C. M. Christianen}
\affiliation{High Field Magnet Laboratory (HFML - EMFL), Radboud University, 6525 ED Nijmegen, The Netherlands}

\author{Christian Sch\"uller}
\affiliation{Institut f\"ur Experimentelle und Angewandte Physik,
	Universit\"at Regensburg, 93040 Regensburg, Germany}

\author{Andreas Knorr}
\affiliation{Institut f\"ur Theoretische Physik, Technische Universit\"at Berlin, 10587 Berlin, Germany}

\author{Tobias Korn\footnote{tobias.korn@uni-rostock.de}}
\affiliation{Institut f\"ur  Physik, Universit\"at Rostock, 18059 Rostock, Germany}
\email{tobias.korn@uni-rostock.de}

\begin{abstract}
In van der Waals heterostructures (HS) consisting of stacked MoSe$_2$ and WSe$_2$ monolayers, optically bright interlayer excitons (ILE) can be observed when the constituent layers are crystallographically aligned. The symmetry of the monolayers allows for two different types of alignment, in which the momentum-direct interlayer transitions are either valley-conserving (R-type alignment) or changing the valley index (H-type anti-alignment). 
Here, we study the valley polarization dynamics of ILE in magnetic fields up to 30~Tesla by time-resolved photoluminescence (PL). For all ILE types, we find a finite initial PL circular degree of polarization ($DoP$) after unpolarized excitation in applied magnetic fields. For ILE in H-type HS, we observe a systematic increase of the PL $DoP$ with time in applied magnetic fields, which saturates at values close to unity for the largest fields. By contrast, for ILE in R-type HS, the PL $DoP$ shows a decrease and a zero crossing before saturating with opposite polarization. This unintuitive behavior can be explained by a model considering the different ILE states in H- and R-type HS and their selection rules coupling PL helicity and valley polarization. 
\end{abstract}

\maketitle
\section{Introduction}
In recent years, two-dimensional (2D) crystals and their van der Waals (vdW) heterostructures~\cite{Geim2013} (HS) have garnered a lot of scientific interest. Besides graphene, the semiconducting transition-metal dichalcogenides (TMDCs) such as MoS$_2$ are probably the most intensely studied 2D crystals. This research activity is motivated by a number of highly interesting properties: in the monolayer limit, their band gap changes from indirect to direct, so that pronounced photoluminescence (PL) can be observed~\cite{Splendiani2010,Mak2010}. Their band structure leads to spin-valley coupling~\cite{Xiao2012}, and the interband selection rules allow for a coupled spin-valley polarization to be prepared and read out via circularly polarized excitation and helicity-resolved PL~\cite{Mak2012}. Alternatively, a valley polarization can be induced using external magnetic fields~\cite{MacNeill2015,Mitioglu2015,Plechinger2016,doi:10.1063/5.0042683}. Combining different TMDC monolayers into van der Waals  heterostructures can lead to a type-II band alignment and corresponding charge separation~\cite{Kosmider13,Tongay13}. The spatially separated electron-hole pairs remain Coulomb-coupled and may form optically bright interlayer excitons (ILE), which have been observed in many TMDC heterobilayer material combinations~\cite{Fang2014,Tongay2014,Xu_NatComm15,Rivera2018}. A novel degree of freedom unique to vdW HS, the interlayer twist angle, plays an important role in governing ILE properties. Depending on the material system, changing the twist angle may tune the ILE emission energy~\cite{Kunstmann18,TEBYETEKERWA2021100509} or control whether ILE are optically bright~\cite{Nayak17}. Among the different material combinations, MoSe$_2$-WSe$_2$ heterobilayers are among the most studied due to the large spectral separation between the ILE and the constituent monolayer exciton emission~\cite{Xu_NatComm15}, the very long photoluminescence lifetimes at low temperatures~\cite{Nagler17,Wurstbauer17} and the large, long-lived valley polarization that can either by induced by circularly polarized excitation~\cite{Xu_Science16} or applied magnetic fields~\cite{Nagler17b}. In this material combination, bright ILE transitions are predominantly momentum-direct and occur between the band extrema at the K points of the constituent monolayers. In order to achieve such momentum-direct transitions, the twist angle can be chosen to be  0 degrees, so that transitions occur between MoSe$_2$ conduction band states and WSe$_2$ valence band states that have the same valley index (R-type or aligned HS). Alternatively, for a twist angle of 60 degrees (H-type or anti-aligned HS), the momentum-direct ILE transitions are between opposite valleys.

More recently, pronounced moiré effects were observed in these HS: when the constituent layers of a vdW HS are not perfectly (anti-)aligned, a moiré pattern develops, whose wavelength strongly depends on the interlayer twist angle. This moiré pattern leads to a periodic potential modulation for ILE in TMDC HS, which may provide potential traps~\cite{Seyler2019,Tran2019}. As a consequence of the moiré periodicity, the local interlayer atomic registry varies, and consequently, the selection rules become spatially dependent~\cite{yu2017moire}. When the interlayer twist angle gets small enough, it becomes energetically favorable for the HS to slightly distort the constituent lattices, so that domains are formed in which there is a well-defined interlayer atomic registry. 
This so-called atomic reconstruction has been observed in recent experiments~\cite{rosenberger2020twist,Weston20,holler2020low}, and regions of atomic reconstruction may  coexist with regions with a moiré-type potential modulation within a single heterostructure sample~\cite{Parzefall_2021}. 

The atomically reconstructed domains can be classified by the specific registry of atoms in the adjacent layers. In the following, we utilize the notation introduced by Yu et al.~\cite{yu2017moire}. For this, three high-symmetry positions within the monolayer lattice are considered, the hollow center of a hexagon (h), a chalcogen atom site (X) and a metal atom site (M). A local atomic registry or a reconstructed domain can then be identified as $R_a^b$, with subscript (a) denoting the hole host layer and superscript (b) denoting the electron host layer of the HS. Thus, for example,  $H_h^M$ corresponds to an anti-aligned (60 degree) structure in which the metal atom site of the electron host layer is above the hollow site of the hole host layer.

Here, we study  ILE  in H-type and R-type MoSe$_2$-WSe$_2$ heterobilayers by helicity- and time-resolved photoluminescence in applied magnetic fields of up to 30~Tesla. In a static, helicity-resolved PL measurement series, we find that ILE in H-type structures have a large, negative g factor, while ILE in R-type structures have a smaller, positive g factor. Time-resolved measurements in applied magnetic fields reveal that in H-type structures, the  PL circular degree of polarization ($DoP$) has a finite positive value immediately after excitation and systematically increases with time, saturating at values close to unity. By contrast, the $DoP$ in R-type structures starts with a positive value, but shows a zero crossing at later times and saturates at negative values. In order to describe these unusual observations, we develop a model to calculate the intra- and interlayer exciton dynamics, taking into account the different interlayer states for H-type and R-type structures. 

\section{Samples and experimental setups}
\subsection{Sample preparation}
Our heterostructures  were fabricated by means of a deterministic transfer process~\cite{Castellanos2014}. For this, we initially exfoliated TMDC flakes from bulk crystals (HQ graphene) onto intermediate polydimethylsiloxane (PDMS) substrates. Monolayer regions of these flakes were identified via optical microscopy. Then, we subsequently transfered the constituent flakes of a heterostructure onto the target substrate, a silicon wafer piece covered with an SiO$_2$ layer and pre-defined metal markers. Well-cleaved, straight edges of the monolayer parts of the two flakes were carefully aligned  to yield crystallographic alignment of the layers.  Subsequent to the transfer, some samples were annealed in vacuum at a temperature of 100$^o$C for several hours to improve interlayer coupling~\cite{Tongay2014}. We note that this preparation method does not allow us to control whether our structures have H-type or R-type alignment. This is determined \textit{a posteriori} using optical spectroscopy, as discussed below.  

\subsection{Optical spectroscopy}
Low-temperature PL characterization measurements were performed in a self-built confocal microscope setup. A frequency-doubled solid-state laser emitting at 532~nm was used for excitation. The laser light was coupled into a 100x microscope objective and focused to a spot diameter of less than 1~micron on the sample surface. The PL from the sample was collected using the same objective and coupled into a grating spectrometer, where it was detected using a charge-coupled device (CCD) sensor. The sample was mounted on the cold finger of a small He-flow cryostat and scanned beneath the microscope objective. 
Low-temperature PL measurements in large magnetic fields were performed at the HFML facility in Nijmegen. The sample was placed on a x-y-z piezoelectric stage and cooled down to 4.2~K in a cryostat filled with liquid helium. Magnetic fields up to 30~T were applied by means of a resistive magnet in Faraday configuration. 
A diode laser (emission wavelength 640~nm) was used for excitation. The laser light was
linearly polarized and focused onto the sample with a microscope objective resulting in a spot size of about 4~micron. The polarization of the PL was analyzed with a quarter-wave plate and a linear polarizer. The PL was then coupled into a grating spectrometer, where it was detected using a CCD sensor. 
For time-resolved measurements,  the diode laser was operated in pulsed mode, with a repetition rate of 1~MHz and a pulse length of about 90~ps. The PL emitted from the sample was filtered with a suitable bandpass and detected with an avalanche photodiode coupled to fast readout electronics. 
\section{Theoretical model}
First, we describe the scope of our model. 
\begin{figure*}[t!]
	\begin{center}
		\includegraphics[width=1.0\linewidth]{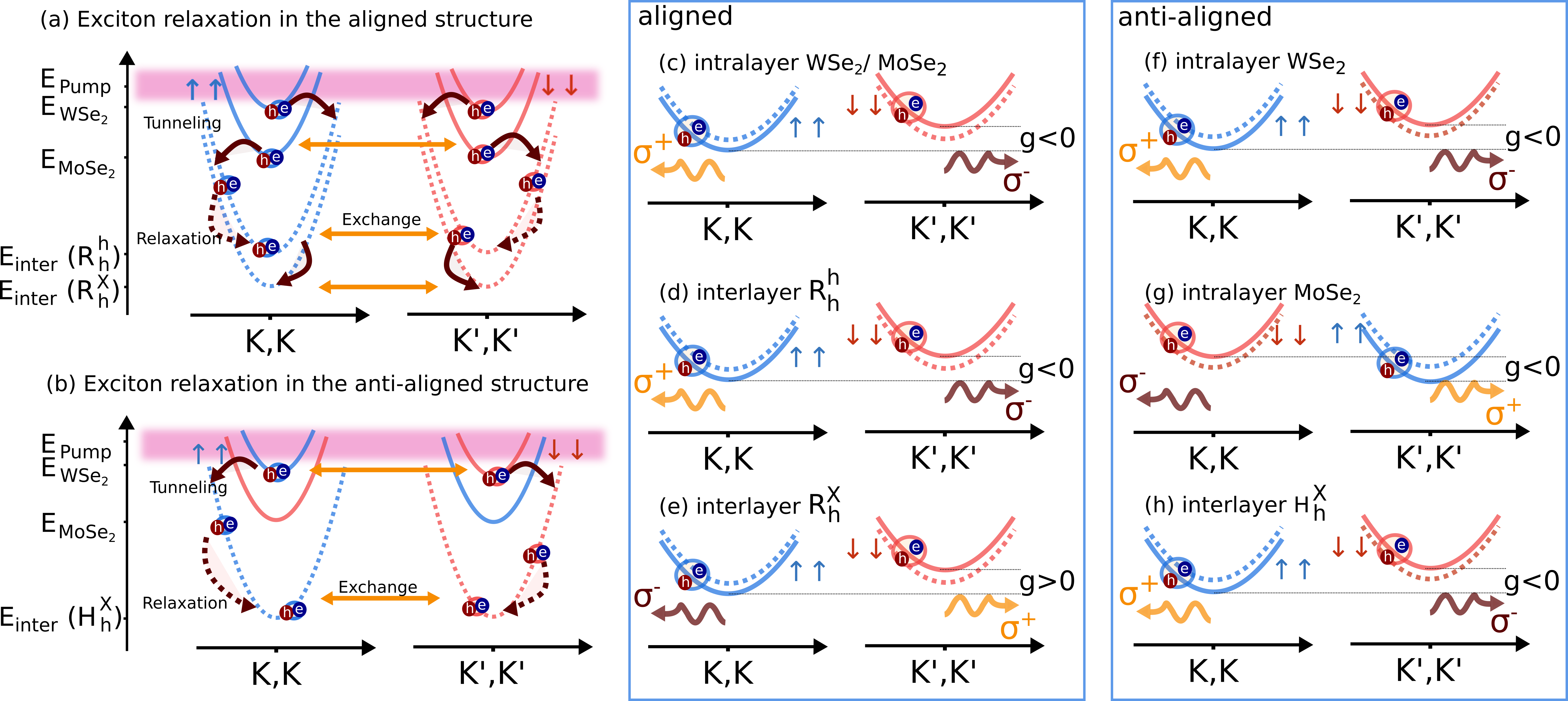}
	\end{center}
	\caption{\textbf{Schematic illustration of the exciton dispersion and g factors} (a) Exciton dispersion and relaxation pathways in the aligned structure. (b) Exciton dispersion and relaxation pathways in the anti-aligned structure . (c-e) g-factors and optical selection rules in the aligned structure of the intralayer excitons (c), the interlayer $R_h^h$ (d) and the interlayer $R_h^X$ exciton (e). (f-h) g-factors and optical selection rules in the anti-aligned structure of the WSe$_2$ intralayer excitons (f), the MoSe$_2$ intralayer exciton (g) and the interlayer $H_h^X$ exciton (h).}
	\label{schema}
\end{figure*}

\textit{Aligned structure:} Figure \ref{schema} (a) illustrates the energy dispersion and relaxation pathways in the aligned structure: the sample is excited slightly above the WSe$_2$ exciton which creates intralayer excitons mainly in WSe$_2$. The $(K,K)$/$(K',K')$ excitons in WSe$_2$ and MoSe$_2$ can be addressed with $\sigma_+$/$\sigma_-$ polarized light and have a g-factor of approximately \unit[-3.7]{}\cite{wozniak2020exciton}, cf. figure \ref{schema} (c). Phonon-assisted tunneling leads to the relaxation of the optically pumped intralayer excitons into $R_h^h$ exciton states~\cite{hagel2021exciton}. These states show similar optical selection rules and a g-factor of approximately \unit[-6.2]{}\cite{wozniak2020exciton}, cf. figure \ref{schema} (d). Subsequently excitons relax  into the $R_h^X$ domain which is located \unit[30]{meV} below the $R_h^h$ exciton states\cite{yu2017moire}. Here, the optical selection rules are inverted w.r.t. the monolayer transitions, resulting  in a positive g-factor of approximately \unit[6.2]{}, cf. figure \ref{schema} (e).

\textit{Anti-aligned structure:} The energy dispersion and the occuring relaxation pathways of the anti-aligned structure are depicted in figure \ref{schema} (b): The optical excitation occurs slightly above the WSe$_2$ transition. Consequently mainly excitons in WSe$_2$ are formed. The $(K,K)$/$(K',K')$ excitons are addressable with $\sigma_+$/$\sigma_-$ light and the g-factor is approximately \unit[-3.6]{}\cite{wozniak2020exciton},cf. figure \ref{schema} (f). Due to the relative rotation with respect to the WSe$_2$ layer, in the MoSe$_2$ layer the optical selection rules are inverted, cf. figure \ref{schema} (g). Phonon-assisted tunneling mediates the relaxation of the optically pumped WSe$_2$ excitons to interlayer $H_h^X$ excitons~\cite{hagel2021exciton}. These excitons have similar optical selection rules as the WSe$_2$ layer and a g-factor of about \unit[-12.6]{}\cite{wozniak2020exciton}, cf. figure \ref{schema} (h).

Starting point for the investigation of the intra- and interlayer exciton dynamics in magnetic fields is the definition of exciton operators\cite{Katsch2018}
\begin{equation}
	P_\mathbf{Q}^{l_h l_e \xi_h \xi_e} = \sum_\mathbf{q} \varphi^{*l_h l_e \xi_h \xi_e}_\mathbf{q} v^{\dagger l_h \xi_h}_{\mathbf{q} - \beta^{l_h l_e \xi_h \xi_e}\mathbf{Q}} c^{l_e \xi_e}_{\mathbf{q} + \alpha^{l_h l_e \xi_h \xi_e}\mathbf{Q}},\label{ex-def}
\end{equation} 
with the center of mass momentum of the exciton $\mathbf{Q}$, the layer indices of hole and electron $l_h$ and $l_e$ and the valley spins of hole and electron $\xi_h$ and $\xi_e$. The valley spin index $\xi$ accounts for a merged quantum number of valley $i$ and spin $s$, i.e. $\xi_{h/e} = (i_{h/e},s_{h/e})$. $\alpha^{l_h l_e \xi_h \xi_e}$ and $\beta^{l_h l_e \xi_h \xi_e}$ account for the relative electron and hole masses $\alpha^{l_h l_e \xi_h \xi_e} = \frac{m_e^{l_e \xi_e}}{m_h^{l_h \xi_h} + m_e^{l_e \xi_e}}$ and $\beta^{l_h l_e \xi_h \xi_e} = \frac{m_h^{l_h \xi_h}}{m_h^{l_h \xi_h}+ m_e^{l_e \xi_e}}$, with electron and hole effective mass $m_e^{l \xi}$ and $m_h^{l \xi}$ being approximated as the masses in the corresponding monolayers  obtained from DFT calculations\cite{Kormanyos2015,ovesen2019interlayer}. The appearing wavefunction $\varphi^{l_h l_e \xi_h \xi_e}_\mathbf{q}$ in eq. \ref{ex-def} is obtained by evaluating the Wannier equation for intra $l_h = l_e$ and interlayer $l_h \neq l_e$ excitons\cite{Ovesen2018} and restricting us to the lowest bound 1s excitons. Besides the wavefunctions, the Wannier equation also gives access to the binding energies of the different exciton states $E_B^{l_hl_e\xi_h\xi_e}$.

The band structure of excitons is then given by
\begin{align}
	E_\mathbf{Q}^{l_hl_e\xi_h\xi_e} &= E_0^{l_hl_e\xi_h\xi_e} + E_B^{l_hl_e\xi_h\xi_e\,1s} + \frac{\hbar^2\mathbf{Q}^2}{2M^{l_hl_e\xi_h\xi_e}}\nonumber \\ &+ \mu_B g^{l_h l_e \xi_h \xi_e} B_z,
\end{align}
where $E_0^{l_hl_e\xi_h\xi_e}$ accounts for the electronic band gap energies, $E_B^{l_hl_e\xi_h\xi_e}$ represents the exciton binding energies and the third term accounts for the kinetic energy of excitons with the excitonic mass $M^{l_hl_e\xi_h\xi_e}$. The last term accounts for Zeeman shifts of the excitons, with Bohrs magneton $\mu_B$ and the excitonic g-factors $g^{l_h l_e \xi_h \xi_e}$, taken from DFT calculations\cite{wozniak2020exciton}. The next step is the parametrization of the excitonic Hamiltonian of the system which accounts for the exciton-phonon interaction\cite{Selig2018,ovesen2019interlayer}, exchange interaction of excitons\cite{selig2019ultrafast} for intra- and interlayer excitons as well as tunneling from intralayer to interlayer exciton states. While tunneling between parallel 2D structures typically conserves the momentum of carriers\cite{lorchat2021excitons}, another scattering process is required to assist the tunneling such that energy and momentum conservation during the tunneling event can be fulfilled. Recent approaches have assumed disordered tunneling barriers giving rise to a relaxed momentum conservation\cite{li2014single,ovesen2019interlayer}. Here however, we do not require this assumption but calculate the momentum relaxation during the tunneling due to exciton-phonon scattering explicitly via a canonical transformation in the Hamiltonian\cite{dong2021observation}, cf. supplementary material section I. 

To calculate the dynamics of excitons we exploit the Heisenberg equation of motion for excitonic transition $\langle P_\mathbf{Q}^{l_h l_e \xi_h \xi_e} \rangle$ and the exciton occupation $N_\mathbf{Q}^{l_h l_e \xi_h \xi_e} = \delta \langle P_\mathbf{Q}^{\dagger l_h l_e \xi_h \xi_e} P_\mathbf{Q}^{l_h l_e \xi_h \xi_e} \rangle$ .

The equation of motion of the optically excitable excitonic transition in the rotating frame reads
\begin{align}
	i \hbar \partial_t \langle \tilde{P}^{l_h l_e\xi_h \xi_e}_\mathbf{0} \rangle&= \left(\Delta^{l_h l_e\xi_h\xi_e}-i \gamma^{l_h l_e\xi_h\xi_e}_\mathbf{0}(\hbar \omega_L)\right) \langle \tilde{P}^{l_h l_e\xi_h\xi_e}_\mathbf{0}\rangle \nonumber \\
	&+ \mathbf{d}^{l_h l_e\xi_h\xi_e}\cdot \mathbf{\tilde{E}}\delta^{\xi_h\xi_e} .\label{transition}
\end{align}
The first term accounts for the oscillation of the excitonic transition detuned to the laser frequency $\Delta^{l_h l_e\xi_h\xi_e} = E_\mathbf{0}^{l_h l_e\xi_h\xi_e} - \hbar \omega_L$ and dephasing $\gamma^{l_h l_e\xi_h\xi_e}_\mathbf{0}(\hbar \omega_L)$ of excitons. The second term accounts for the optical source with the dipole moment $\mathbf{d}^{l_h l_e\xi_h\xi_e}$ and the envelope of the exciting field $\tilde{\mathbf{E}}$. The appearing delta function accounts for the conservation of spin and momentum during the optical excitation. As a result, only excitons where the hole and electron valley-spin coincide are optically adressable, i.e. $\xi_h = \xi_e$. Given that interlayer excitons have very small dipole elements compared to intralayer excitons\cite{wozniak2020exciton}, we consider the excitonic polarization for intralayer excitons only.
The equation of motion of the incoherent exciton occupation reads
\begin{align}
	\partial_t N_\mathbf{Q}^{l_h l_e \xi_h\xi_e}&= \sum_{\xi_e'} \Gamma^{in\,l_h l_e \xi_e \xi_h , \xi_e' }_\mathbf{Q} |\langle P^{l_h l_e \xi_h\xi_e'}_\mathbf{0}\rangle|^2 \delta_{\xi_h,\xi_e'} \nonumber \\ 
	&+\sum_{\mathbf{K},\xi_e'} \Gamma_\mathbf{Q,K}^{in\,l_h l_e \xi_e-\xi_e'} N^{l_h l_e \xi_h\xi_e'}_\mathbf{K} \nonumber \\ &- \sum_{\mathbf{K},\xi_e'} \Gamma_\mathbf{Q,K}^{out\,l_h l_e \xi_e - \xi_e'} N^{l_h l_e \xi_h \xi_e}_\mathbf{Q} \nonumber \\&- \Gamma^{l_h l_e \xi_h \xi_e}_{\text{rad}\,\mathbf{Q}} N^{l_h l_e \xi_h\xi_e}_\mathbf{Q} \delta_{\xi_h,\xi_e} \nonumber \\ 
	&+\sum_{\mathbf{K},\xi_e',l_h',l_e'} \Gamma_{T,\mathbf{Q,K}}^{in\,l_h l_e l_h' l_e' \xi_e-\xi_e'} N^{l_h' l_e' \xi_h\xi_e'}_\mathbf{K} \nonumber \\ 
	&- \sum_{\mathbf{K},\xi_e'} \Gamma_{T,\mathbf{Q,K}}^{out\,l_h l_e \xi_e - \xi_e'} N^{l_h l_e \xi_h \xi_e}_\mathbf{Q}.\nonumber \\
	&+\frac{2}{\hbar} \Im\left( X^{l_h  l_e \xi_h \bar{\xi}_h}_\mathbf{Q} C^{l_h l_e \xi_h \bar{\xi}_h}_\mathbf{Q} \delta_{\xi_h, \xi_e} \right)  \label{Density} 
\end{align}

The first line accounts for the formation of incoherent intralayer excitons from coherent intralayer excitons. The second and third line represent the thermalization of intra- and interlayer excitons through exciton-phonon scattering\cite{ovesen2019interlayer,Selig2018}. The fourth line accounts for radiative decay of intra- and interlayer excitons within the radiative cone\cite{ovesen2019interlayer}. All these couplings are diagonal in the layer quantum number and do not provide a transfer between intra- and interlayer states. The fifth and sixth line account for the tunneling of the carriers between the layers leading to coupling between intra- and interlayer excitons, where the first term accounts for inscattering and the second term for outscattering. The tunneling is discussed below in detail. The last line in eq. \ref{Density} accounts for intervalley exchange coupling of intra- and interlayer excitons.  The appearing Kronecker delta accounts for the fact, that only intravalley excitons contribute to the exchange coupling. The interexcitonic transition $C^{l_h l_e \xi \bar{\xi}}_{\mathbf{Q}} = \delta \langle P^{\dagger l_h l_e \xi \xi}_\mathbf{Q} P^{l_h l_e \bar{\xi} \bar{\xi}}_\mathbf{Q} \rangle$ mediates the exchange coupling. 

Its equation of motion reads

\begin{align}
	\partial_t C^{l_h l_e \xi \bar{\xi}}_{\mathbf{Q}} & = \frac{1}{i\hbar}\Delta E^{l_h l_e \xi \bar{\xi}}_\mathbf{Q} C^{l_h l_e \xi \bar{\xi}}_{\mathbf{Q}} \nonumber \\ &+ \frac{1}{i\hbar}X^{l_h  l_e \bar{\xi} \xi }_\mathbf{Q} \left( e^{\Delta_C} N_\mathbf{Q}^{l_h l_e \xi \xi} -e^{-\Delta_C } N_\mathbf{Q}^{l_h l_e \bar{\xi}\bar{\xi}}\right)\nonumber \\
	&-\sum_{\mathbf{K},\xi_e} \frac{1}{2}\left(\Gamma_\mathbf{Q,K}^{out\,l_h l_e \xi - \xi_e} +\Gamma_\mathbf{Q,K}^{out\,l_h l_e \bar{\xi} - \xi_e} \right) C^{l_h l_e \xi \bar{\xi}}_{\mathbf{Q}}.  \label{ccc}
\end{align}

The first term accounts for the oscillation of the interexcitonic transition with the energy difference of the states in different valleys, $\Delta_C =\left( E_\mathbf{Q}^{l_h l_e \xi \xi } - E_\mathbf{Q}^{l_h l_e \bar{\xi} \bar{\xi}} \right) / \left( 2 k_B T \right) $. The second term accounts for the source of the interexcitonic transition. Without magnetic field, i.e.$ \Delta E^{l_h l_e \xi \bar{\xi}}_\mathbf{Q} = 0$, it is driven by the occupation difference of opposite valleys.

 We have added the appearing exponential factors phenomenologically to account for a detailed balance of the exchange coupling in the presence of magnetic fields\cite{allan2007energy}. The third line accounts for exciton-phonon interaction of the interexcitonic transition which leads to a damping of the latter. All coupling elements are defined in the supplementary material section I.

The intensity of the emitted light of the interlayer region reads
\begin{align}
	I^{\sigma}\propto\frac{2 \pi}{\hbar}\hspace{-2pt}\sum_{{\mathbf{K},K_z},\xi} |d^{l_h l_e \xi\xi \sigma}_{\mathbf{K}}|^2  N_\mathbf{K}^{l_h l_e \xi\xi}\delta( \Delta E^{l_h l_e\xi \sigma}_{\mathbf{K},K_z} ).\label{eqPL}
\end{align}
and is determined by the amount of excitons which is located inside of the radiative cone\cite{Selig2018}. The appearing delta function accounts for the conservation of energy and momentum during the photoemission event, $\Delta E^{l_h l_e\xi \sigma}_{\mathbf{K},K_z} = E^{l_h l_e\xi \xi}_{\mathbf{K}} - \hbar \omega^{ \sigma}_{\mathbf{K},K_z} $, with $\hbar \omega^{ \sigma}_{\mathbf{K},K_z} $ being the photon dispersion with the polarization $\sigma$ and the photon momentum in-plane and out-of plane momentum $\mathbf{K}$ and $K_z$.

Atomic reconstruction leads to the formation of different domains for the interlayer excitons\cite{rosenberger2020twist,wozniak2020exciton,holler2020low}. The domains of the heterostructure are treated with a mesoscopic model, assuming that each domain is smeared out homogeneously over the structure. This corresponds to a spatial averaging of the optical excitation. For the aligned structure, i.e. stacking angle of $\theta\approx$\unit[0]{$^\circ$}, we assume that the tunneling from intra- to interlayer excitons occurs solely to the $R_h^h$ domains since here the atoms of both layers are placed directly on top of each other\cite{wozniak2020exciton}. The $R_h^X$ states are located about \unit[30]{meV} below the $R_h^h$ states\cite{yu2017moire}. The coupling between the $R_h^h$ and the $R_h^X$ structure is assumed to be momentum-  and spin-conserving. To account for the coupling between the domains, we include an effective transfer rate to the equations of motion of the interlayer excitons

\begin{align}
	&\partial_t N_\mathbf{Q}^{l_hl_e\xi_h\xi_e} (R_h^h)\Bigg\vert_{dc} = \nonumber \\ 
	&= \chi \left(e^{-\Delta_D } N_\mathbf{Q}^{l_hl_e\xi_h\xi_e} (R_h^X) - e^{\Delta_D } N_\mathbf{Q}^{l_hl_e\xi_h\xi_e} (R_h^h) \right) \label{eom_domains}
\end{align}

where we adjust the constant $\chi$ to the experiment. The exponential factors which depend on the energetic splitting between the domains $\Delta_D = \left( E^{l_hl_e\xi_h\xi_e}_\mathbf{Q}(R_h^h) - E^{l_hl_e\xi_h\xi_e}_\mathbf{Q}(R_h^X) \right) / \left( 2 k_B T \right)$ are added phenomenologically to account for the detailed balance of the relaxation between the different domains. We omit the coupling to the $R^M_h$ structure, since the tunnel coupling from the intralayer monolayer states is assumed to be weak due to the mismatch of the transition metal atoms which host valence and conduction bands. Furthermore the $R^M_h$
states are located energetically above the $R^h_h$ states by more than \unit[50]{meV}\cite{yu2017moire} and have vanishing oscillator strength\cite{yu2017moire,wozniak2020exciton}.  The g factors and oscillator strengths for the $R^h_h$ and $R^X_h$ interlayer excitons are taken from DFT calculations\cite{wozniak2020exciton}.

For the anti-aligned structure, i.e. stacking angle of $\theta\approx$\unit[60]{$^\circ$}, we assume that the tunneling from intra- to interlayer excitons occurs solely to the $H_h^X$ domain, since here the transition metal atoms, , which carry the electronic bands, are placed on top of each other. The $H_h^M$ states are located below the $H_h^X$ by \unit[10]{meV}\cite{yu2017moire}, but since the $H_h^X$ has vanishing oscillator strength \cite{yu2017moire,wozniak2020exciton} we omit the coupling among these two.
Also we omit the coupling to the $H_h^h$ since we assume the tunneling to be weak due to the spatial mismatch of the transition metal atoms. Furthermore it is located about \unit[20]{meV} above the $H_h^X$ and can therefore assumed not to be prominently occupied. The g factors and oscillator strengths for the $H^X_h$ interlayer excitons are taken from DFT calculations\cite{wozniak2020exciton}.
\section{Experimental results}
First, we discuss the spectral characteristics of the PL emission from our heterostructures, which we utilize to determine their alignment.
\begin{figure}
	\begin{center}
		\includegraphics[width=1.0\linewidth]{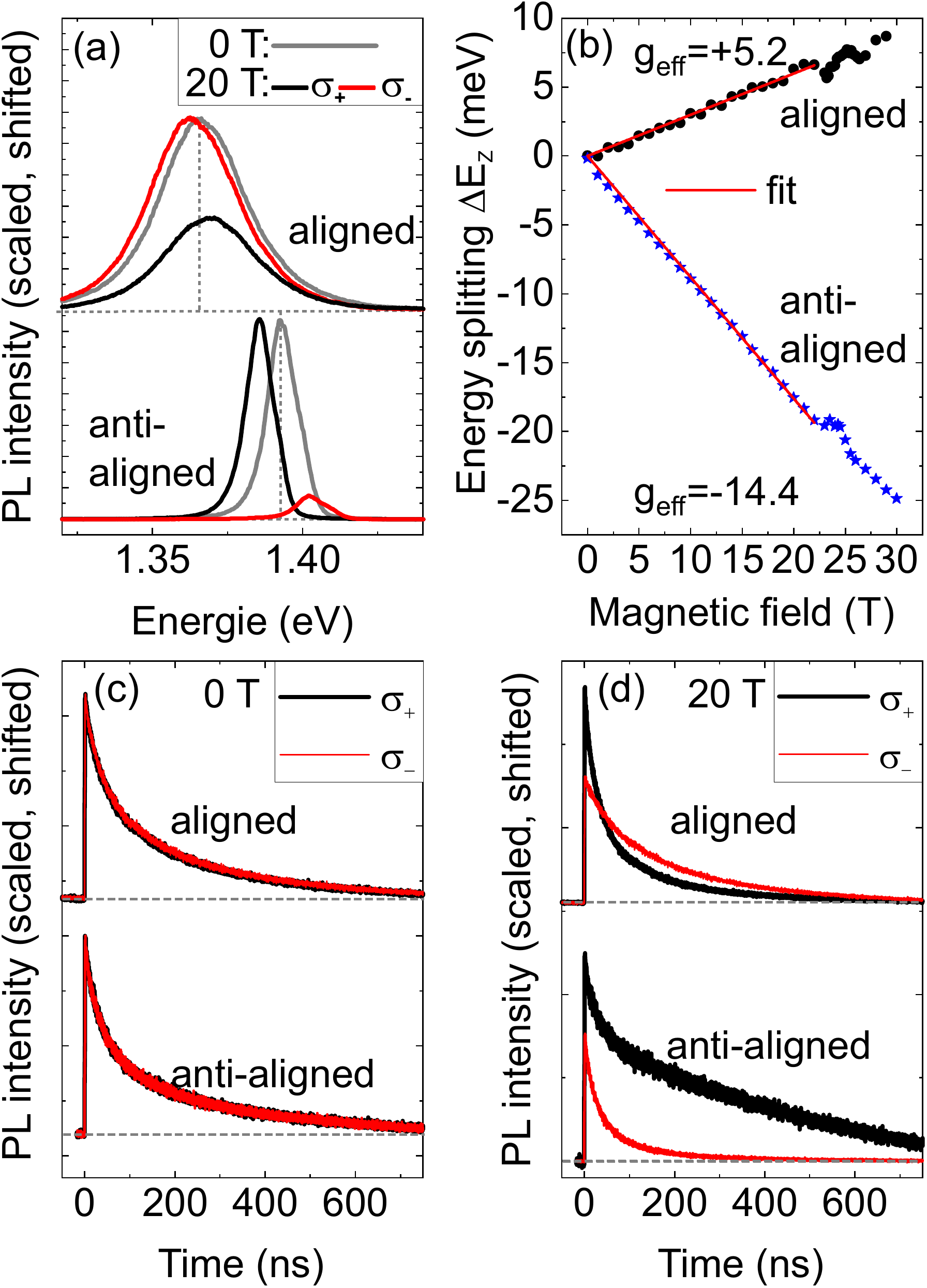}
	\end{center}
	\caption{(a) PL spectra of aligned and anti-aligned HS measured without applied magnetic field (grey lines) and helicity-resolved spectra at 20~T applied magnetic field (black and red lines). The vertical dashed lines indicate the peak positions at zero magnetic field and serve as a guide to the eye. (b) Magnetic-field-dependent energy splitting $\Delta E_Z$ of ILE in aligned (black dots) and anti-aligned (blue stars) HS. The solid lines indicate linear fits to the data, which were used to exctract the effective g factors. (c-d) Helicity-resolved ILE PL traces for aligned and anti-aligned HS at 0~T (c) and 20~T (d).}
	\label{Charakt-4Panel}
\end{figure}
Figure~\ref{Charakt-4Panel}(a) shows PL spectra of the two HS investigated in this study in the spectral region of the ILE. Let us first compare the spectra measured without applied magnetic field (grey lines), which have been equally scaled to allow for easy comparison.  We clearly see a pronounced difference between the two HS: while one sample shows an ILE emission centered around 1367~meV with a FWHM of about 36~meV, the other sample has a higher-energy ILE emission centered at 1393~meV with a much narrower FWHM of only about 13~meV. These two types of emission energies and linewidths are characteristic for aligned (low energy, larger linewidth) and anti-aligned (high energy, smaller linewidth) HS, respectively, as we could verify using HS containing, both, aligned and anti-aligned regions fabricated in a stack-and-tear process~\cite{Holler-diamagnetic}. 

We further confirm this assignment using magnetic-field-dependent PL spectra. Under linearly polarized excitation, the $\sigma_+$ and $\sigma_-$ components of the emission have the same energy and intensity at 0~T. In an applied magnetic field, however, a helicity-dependent energy splitting and preferred emission helicity develop, as seen in Fig.~\ref{Charakt-4Panel}(a). We note that the finite-field spectra have been scaled to reflect their relative intensities. We clearly see that for both HS, the helicity component emitting at lower energy becomes more intense than the higher-energy emission. Correspondingly, a circular degree of polarization ($DoP$) can be defined as 
\begin{equation}\label{Dop}
DoP = \frac{I^{\sigma_+}-I^{\sigma_-}}{I^{\sigma_+}+I^{\sigma_-}}  
\end{equation}
with the PL intensities $I^{\sigma_+}$ and $I^{\sigma_-}$ corresponding to the two different emission helicities. Based on this definition, we see that the $DoP$ is positive for the anti-aligned HS, as its $\sigma_+$ component shifts to lower energy and becomes more intense than the $\sigma_-$ component,  and negative for the aligned HS, where the two components behave in the opposite way in  continuous-wave PL measurements. 
The Zeeman-like helicity-dependent energy splitting $\Delta E_Z$ is linear in the magnetic field and can be described by
\begin{equation}\label{gfactor}
	\Delta E_Z(B)=E^{\sigma_+}(B)-E^{\sigma_-}(B) \equiv g_{eff} \mu_{B} B
\end{equation} 
with the Bohr magneton $\mu_{B}$ and the effective Landé g factor $g_{eff}$, whose sign is defined based on the emission helicities. By extracting the peak positions of the $\sigma_+$ and $\sigma_-$ components from PL spectra measured in different magnetic fields using Gaussian peak fitting, we can extract $\Delta E_Z(B)$ for our two samples and determine $g_{eff}$ from a linear fit to the data, as shown in Fig.~\ref{Charakt-4Panel}(b).
We restrict the fit range from 0 to 20~T, as for both samples, a pronounced nonlinearity of the energy splitting can observed around 24~T. This feature has already been observed in magneto-PL studies of anti-aligned HS by some of the authors~\cite{Nagler17b} and more recently been associated with exciton-phonon interaction~\cite{Delhomme_2020}. 
We find $g_{eff}=+5.2$ for the aligned and $g_{eff}=-14.4$ for the anti-aligned HS, in qualitative agreement with recent studies~\cite{Nagler17b,Sey19,Ciarrocchi2019,Delhomme_2020}. 

Next, we turn to time-resolved PL measurements. Figure~\ref{Charakt-4Panel}(c) shows time traces of ILE PL emission for aligned and anti-aligned HS in the absence of a magnetic field. As in the continuous-wave measurements discussed above, linearly polarized excitation yields similar intensities and dynamics for the $\sigma_+$- and $\sigma_-$-polarized emission. After pulsed excitation, the emission in both HS monotonously decays over a time window of several hundred nanoseconds. The decay can be well-described with a biexponential fit yielding similar decay constants of $t_1=37$~ns and $t_2=240$~ns for the aligned HS and $t_1=37$~ns and $t_2=286$~ns for the anti-aligned HS, respectively. 

Remarkably, in an applied magnetic field, the helicity-resolved dynamics of aligned and anti-aligned HS differ substantially, as Fig.~\ref{Charakt-4Panel}(d) demonstrates. In the anti-aligned structure, the intensity of the $\sigma_+$ emission is larger than that of the $\sigma_-$ emission immediately after excitation, and remains so throughout the whole time window. We also find that the PL lifetime for $\sigma_+$ becomes longer, while it becomes shorter for $\sigma_-$. By contrast, in the aligned structure, while the $\sigma_+$ emission is initially larger than the $\sigma_-$ component, its lifetime is shorter, so that at a finite time after excitation, the $\sigma_-$ component becomes larger. 
\begin{figure}[h!]
	\begin{center}
		\includegraphics[width=1.0\linewidth]{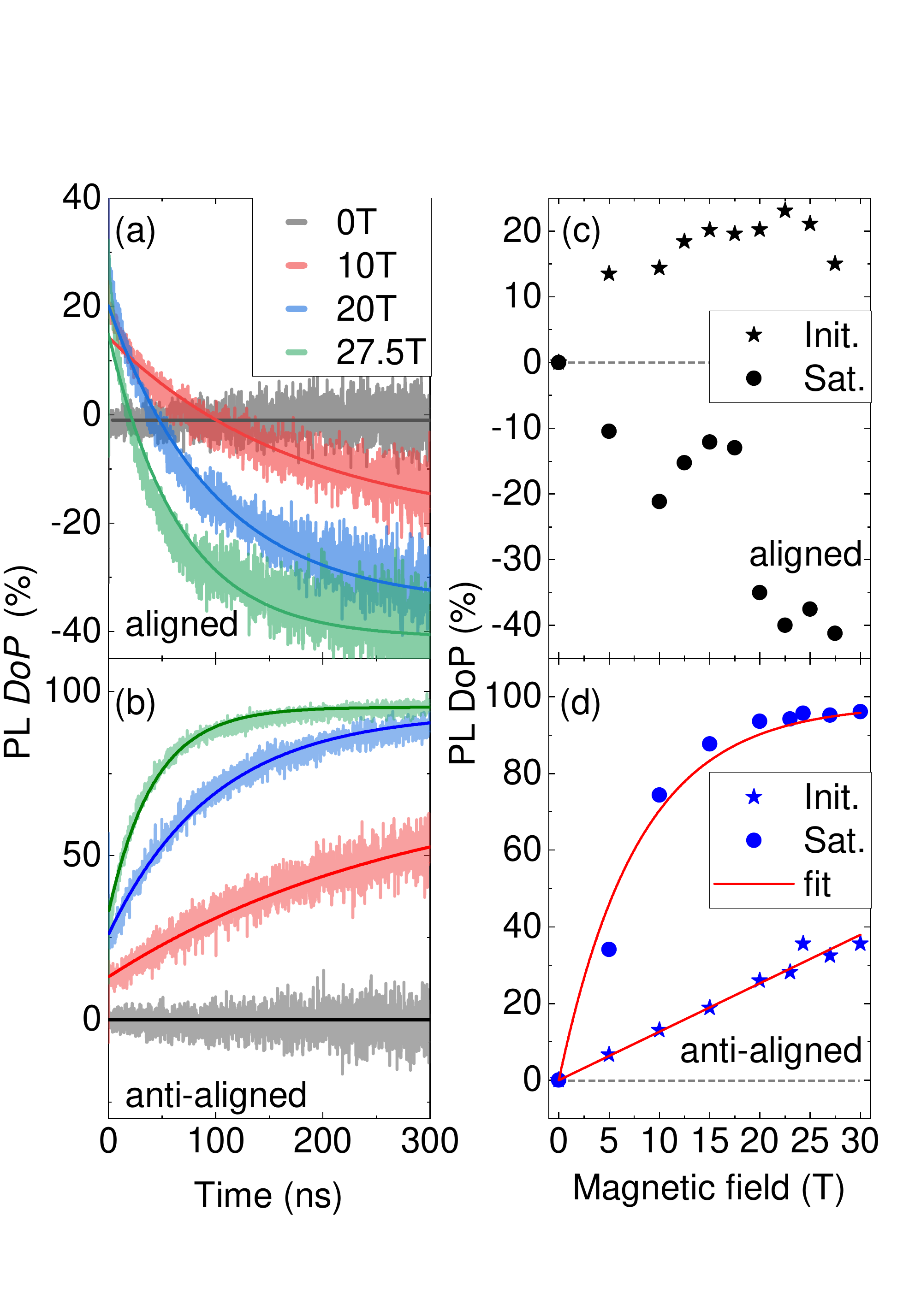}
	\end{center}
	\caption{\textbf{Degree of Polarization of the Photoluminescence} in the (a) aligned and (b) anti-aligned structure as a funtion of time for various magnetic fields. The semi-transparent lines represent the data calculated from helicity-resolved PL traces, the solid lines are fits to the data. (c+d) initial (stars) and saturation (dots) values of the $DoP$ for aligned (c) and anti-aligned (d) HS extracted from the data in (a+b).}
	\label{PolDeg-Exp-4Panel-hochkant}
\end{figure}

As a consequence,  the $DoP$ for aligned and anti-aligned HS shows very different dynamics in applied magnetic fields, as can be seen in Fig.~\ref{PolDeg-Exp-4Panel-hochkant}: panels (a) and (b) show the time-resolved $DoP$ calculated from helicity-resolved TRPL traces. We first discuss its dynamics in the anti-aligned structure (Fig.~\ref{PolDeg-Exp-4Panel-hochkant}(b)). Here, the $DoP$ has a finite, positive value immediately after excitation, which monotonically increases as a function of the applied field. At later times, the $DoP$ increases from its start value and reaches a saturation level.   With increasing magnetic fields, the maximum $DoP$ becomes larger, and saturation is reached faster. The $DoP$ dynamics can be well-described using a saturation curve:
\begin{equation}\label{Dop-t}
	DoP(t) = P_f+(P_i-P_f)e^{-\frac{t}{\tau}}
\end{equation}
with  $P_i$ the initial and $P_f$ the saturation value of the  $DoP$ and the relaxation time $\tau$.
Initial and saturation values extracted from the data are depicted in Fig.~\ref{PolDeg-Exp-4Panel-hochkant}(d). Here, we find that the dependence of the initial $DoP$ on the magnetic field can be well-described by a linear fit, while the saturation value, which reaches almost 100~percent at high fields, can be described by an exponential saturation curve $DoP(B)=(1-e^{-\frac{B}{B_{Sat}}})*P_{Sat}$ with a characteristic field $B_{Sat}\approx8$~T.  
By contrast, the aligned structure shows a very different behavior, as seen in  Fig.~\ref{PolDeg-Exp-4Panel-hochkant}(a). Immediately after excitation, the $DoP$ has a finite, positive value. However, as a function of time, the $DoP$ decreases, crosses the zero line and saturates at a negative value. With increasing magnetic field, the zero crossing occurs at earlier times, the negative saturation value increases, and the saturation level is reached faster. We also fitted this behavior  based on equation~\ref{Dop-t}, with opposite sign of $P_i$ and $P_f$. However, this fit function systematically yields $P_i$ values that are too low. 
Neither the initial nor the saturation $DoP$ values can be described by simple fit functions for the aligned HS, as Fig.~\ref{PolDeg-Exp-4Panel-hochkant}(c) shows.  
In order to understand the  $DoP$ dynamics in our HS, particularly the unusual behavior in the aligned structure, we turn to our theoretical model of the heterostructures.

\section{Theory results and discussion}
\begin{figure}[h!]
	\begin{center}
		\includegraphics[width=1.0\linewidth]{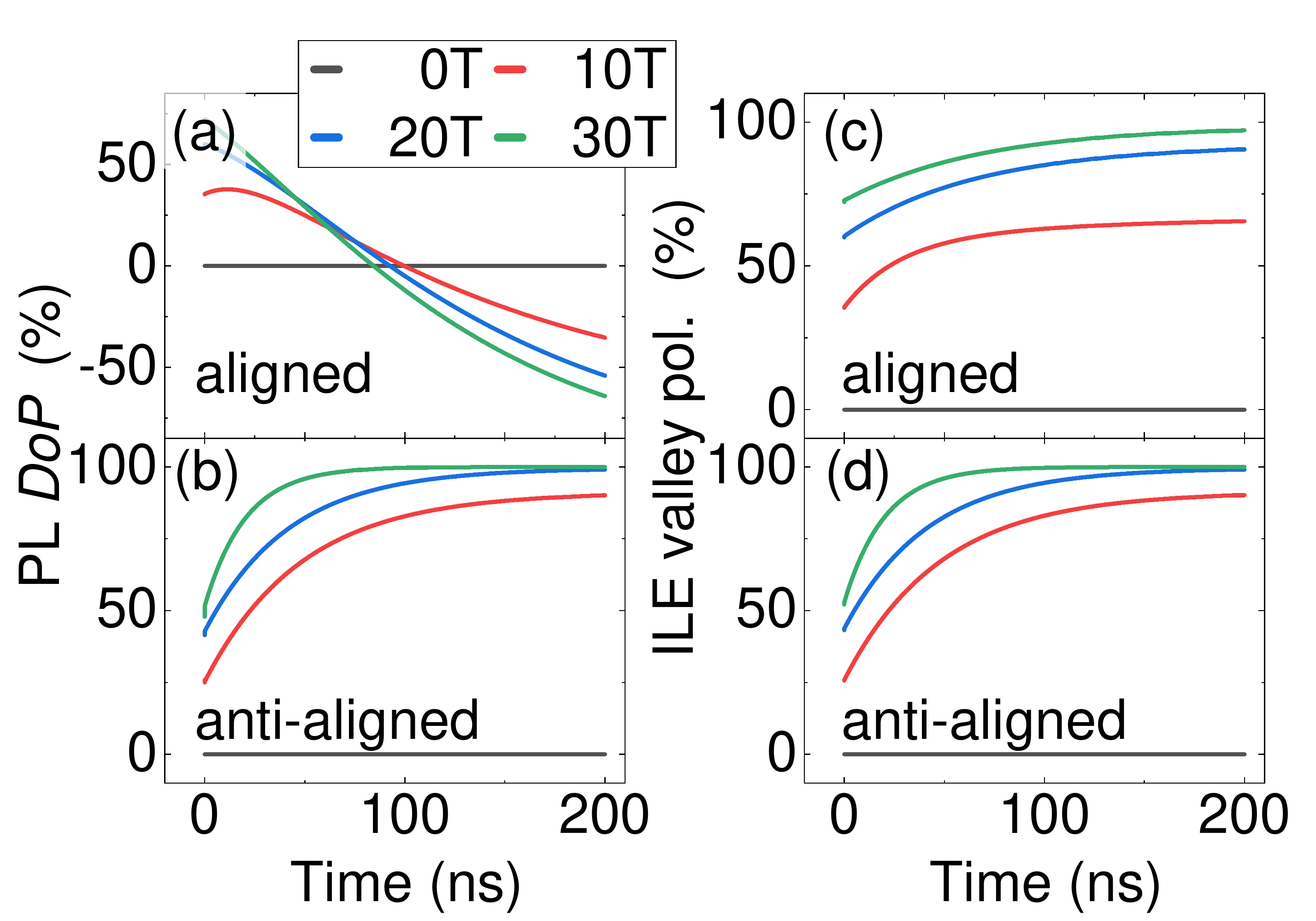}
	\end{center}
	\caption{Degree of Polarization of the Photoluminescence in the (a) aligned and (b) anti-aligned structure as a funtion of time for various magnetic fields. Interlayer exciton valley polarization degree in the (c) aligned and (d) anti-aligned structure as a funtion of time for various magnetic fields.}
	\label{PolDeg-Expol-4Panel}
\end{figure}
Using our model, we can calculate the $DoP$ dynamics and compare the results to the experiments.   
First, we investigate the anti-aligned structure (Figure~\ref{PolDeg-Expol-4Panel}(b)).  We find that the photoluminescence is already polarized at t=0, i.e., during the formation of ILE. The initial polarization increases as a function of the applied magnetic field. We attribute the PL from the interlayer exciton to the $H_h^X$ states which are formed through phonon-assisted tunneling from the optically pumped WSe$_2$ intralayer excitons. The applied magnetic field leads to a polarization of the WSe$_2$ intralayer excitons which directly translates into the polarization of interlayer excitons during the tunneling and thus explains the observed initial $DoP$. Microscopically, the initial $DoP$ arises  from the interplay of exchange coupling in the intralayer states and phonon-assisted tunneling between intra- and interlayer states, see equation~\ref{Density}. The exchange coupling mainly depends on the dipole moment of the intralayer excitons, taken from DFT calculations~\cite{Xiao2012}. The tunneling strength is given mainly by the wave function overlap between both layers~\cite{ovesen2019interlayer}, taken from DFT calculations, and the exciton-phonon  coupling strength~\cite{Selig2018}, also parametrized from DFT calculations~\cite{PhysRevB.90.045422}.

 The linear dependence of the initial $DoP$ on magnetic field observed in experiment (Fig.~\ref{PolDeg-Exp-4Panel-hochkant}(d)) indicates that tunneling from intralayer to interlayer states is too fast to allow a quasi-equilibrium of the WSe$_2$ intralayer exciton valley polarization to develop. 

In the course of time, we find a further increase of the DoP, which is due to the fact that the $H_h^X$ interlayer excitons have a larger $g$ factor compared to $WSe_2$ intralayer excitons, i.e. the energetic splitting of $\uparrow\uparrow$ and $\downarrow\downarrow$ excitons is larger in the interlayer state compared to the WSe$_2$ intralayer states, which results in a larger occupation difference in the thermal limit and thus a larger degree of polarization. The calcaluted $DoP$ dynamics qualitatively matches the experimental behavior in the anti-aligned HS. Microscopically, the  observed timescale is dictated by the intervalley exchange coupling of interlayer excitons (equation~\ref{Density}), which depends on the dipole moment of the interlayer excitons which are taken from DFT calculations~\cite{wozniak2020exciton}.

Next, we turn to the $DoP$ dynamics in the aligned structure depicted in Fig.~\ref{PolDeg-Expol-4Panel}(a).  Again, we find that the ILE emission is already polarized immediately after the optical pump of the system. This initial polarization increases as a function of the applied magnetic field. We assign the positive degree of polarization to the emission from excitons in the $R_h^h$ domain, which are dominantly formed through phonon-assisted tunneling directly from the WSe$_2$ intralayer excitons. Since exchange coupling and phonon-assisted tunneling occur on similar timescales, this intralayer polarization translates directly into polarization of the interlayer excitons in the $R_h^h$ domain. Similar to the anti-aligned structure, the observed initial polarization is determined via the strength of the exchange coupling in the monolayer states and the efficiency of the phonon-assisted tunneling.
Given that both species, the WSe$_2$ intralayer excitons as well as the interlayer excitons in the $R_h^h$ exhibit a negative g factor and share the same optical selection rules, the initial polarization of the interlayer excitons is positive.

In the course of time, the $DoP$ decreases until it changes the sign and starts to saturate at the opposite $DoP$.  We assign this sign change to a transition of the excitons from the $R_h^h$ domain to the $R_h^X$. In the $R_h^X$ domain, the valley optical selection rules as well as the $g$ factors are inverted with respect to the $R_h^h$ domain as sketched in Fig.~\ref{schema} (d),(e). This results in a crossover of the degree of polarization. Interestingly, this crossover is not accompanied by a spin flip of the excitons, since in both, the $R_h^h$ and the $R_h^X$ domain, the $\uparrow\uparrow$ excitons are energetically below the $\downarrow\downarrow$ for a positive magnetic field, cf. \ref{schema} (d),(e). 
The transition rate of  $R^h_h$ to $R_h^X$ excitons depends on the phenomenologically chosen parameter $\chi$ in equation~\ref{eom_domains}.

Futhermore, we observe that the zero crossing moves towards shorter times as the magnetic field increases. This observation is asigned to the coupling between the $R_h^h$ and the $R_h^X$ domain. While the absolute value of the $g$ factor in the $R^X_h$ domain is smaller compared to the value in the $R_h^h$ domain, the relative splitting of $\uparrow\uparrow$ states decreases in comparison to the splitting of the $\downarrow\downarrow$ states in both domains which favors the interdomain relaxation of $\uparrow\uparrow$ excitons. 

Thus, we can understand the  behavior of the $DoP$ in both, H-type and R-type HS to be driven by an energy relaxation of ILE towards the lowest-energy states. The dynamics in H-type HS is rather straightforward, as there is a single dominant interlayer exciton state, $H_h^X$, which has the same selection rules coupling valley index and light helicity as the intralayer state. In R-type HS, the situation is more complex, as there are two relevant interlayer exciton states, $R_h^h$ and  $R_h^X$ with opposite selection rules, and the observed {$DoP$} arises from a superposition of emission from these states. We stress that in both, H-type and R-type HS, the ILE valley polarization degree systematically \textit{increases} as a function of time. This is clearly seen in Fig.~\ref{PolDeg-Expol-4Panel} (c) and (d). For the anti-aligned HS, we see that the ILE valley polarization degree matches the PL {$DoP$}. By contrast, for the aligned HS, the PL {$DoP$} shows the peculiar zero crossing behavior discussed above, while the ILE valley polarization degree has a qualitatively similar behavior as for the anti-aligned structure. As the effective g factor in the aligned structure is smaller, saturation of the ILE valley polarization degree is reached at later times and higher fields than in the anti-aligned structure.

\section{Conclusions}
In summary, we have studied the valley polarization dynamics of interlayer excitons in MoSe$_2$-WSe$_2$ heterobilayers in large applied magnetic fields using time-resolved photoluminescence. Our experimental findings are well-reproduced using a model taking into account  intra- and interlayer exciton dynamics and the different variety of interlayer states in R- and H-type heterostructures.   We find that in finite applied magnetic fields, a substantial initial circular polarization degree of the photoluminescence can be observed. We associate this with a magnetic-field-induced valley polarization of intralayer excitons that occurs even before interlayer excitons are formed via interlayer charge transfer. The subsequent dynamics of the circular polarization depends on the heterostructure alignment. In H-type heterostructures, the circular polarization degree systematically grows and saturates as a function of time. By contrast, in R-type heterostructures, it decreases and saturates with opposite sign after a zero crossing.   We can understand this unusual behavior by considering the optically bright interlayer exciton states for the different alignment types, their effective g factors and their respective selection rules coupling the valley polarization to the light helicity. 

\acknowledgements 
We gratefully acknowledge financial support by the DFG via the following projects: GRK 1570 (J. H., P.N., C.S.), KO3612/3-1(project-ID 631210), KO3612/4-1(project-ID 648265) (T.K.), SFB1277 (project B05, T.K., M.K., A.C., C.S.), Emmy-Noether Programme (CH 1672/1, A.C.), W\"urzburg-Dresden Cluster of Excellence on Complexity and Topology in Quantum Matter ct.qmat (EXC 2147, Project-ID 390858490, A.C.),  Walter-Benjamin Programme (project-ID 462503440) (J.Z.) SFB 951 (Project-ID 182087777, project B12, M.S., M.K., A.K.) and KN 427/11-1 (Project-ID 420760124, F.K. and A.K.)  
This work was supported by HFML-RU/NWO-I, member of the European Magnetic Field Laboratory (EMFL).

J.H. and M.S.  contributed equally to this work.

\end{document}